\documentclass[twocolumn,aps,prx,10pt,superscriptaddress]{revtex4-1}
\usepackage{times}
\usepackage{amsmath}    
\usepackage{graphicx}   
\usepackage{verbatim}   
\usepackage{color}      
\usepackage{hyperref}   
\usepackage{bbold}

\usepackage{textcomp}
\usepackage{lpic}

\usepackage[normalem]{ulem}

\newcommand{\Cliff}{\mathsf{G}}
\newcommand{\cC}{\mathcal{C}}
\newcommand{\cD}{\mathcal{D}}
\newcommand{\cE}{\mathcal{E}}
\newcommand{\cF}{\mathcal{F}}

\newcommand{\Tr}{\mathrm{Tr}}
\newcommand{\ket}[1]{\lvert#1\rangle}
\newcommand{\bra}[1]{\langle #1\rvert}
\newcommand{\proj}[1]{\ket{#1}\!\bra{#1}}
\newcommand{\Fup}{\bar{F}_m^\uparrow}
\newcommand{\Fdown}{\bar{F}_m^\downarrow}

\definecolor{orange}{rgb}{1,0.5,0}

\begin{document}

\title{Non-exponential Fidelity Decay in Randomized Benchmarking with Low-Frequency Noise}

\author{M.~A.\ Fogarty}
\author{M.\ Veldhorst}
\affiliation{Centre for Quantum Computation and Communication Technology, 
School of Electrical Engineering and Telecommunications, 
The University of New South Wales, Sydney, NSW 2052, Australia.}

\author{R.\ Harper}
\affiliation{Centre for Engineered Quantum Systems, School of Physics, The University of Sydney, Sydney, NSW 2006, Australia}

\author{C.~H.\ Yang}
\affiliation{Centre for Quantum Computation and Communication Technology, 
School of Electrical Engineering and Telecommunications, 
The University of New South Wales, Sydney, NSW 2052, Australia.}

\author{S.~D.\ Bartlett}
\author{S.~T.\ Flammia}
\affiliation{Centre for Engineered Quantum Systems, School of Physics, The University of Sydney, Sydney, NSW 2006, Australia}

\author{A.~S.\ Dzurak}
\affiliation{Centre for Quantum Computation and Communication Technology, 
School of Electrical Engineering and Telecommunications, 
The University of New South Wales, Sydney, NSW 2052, Australia.}

\date{\today}

\begin{abstract}
We show that non-exponential fidelity decays in randomized benchmarking experiments on quantum dot qubits are consistent with numerical simulations that incorporate low-frequency noise. By expanding standard randomized benchmarking analysis to this experimental regime, we find that such non-exponential decays are better modeled by multiple exponential decay rates, leading to an \textit{instantaneous control fidelity} for isotopically-purified-silicon MOS quantum dot qubits which can be as high as 99.9\% when low-frequency noise conditions and system calibrations are favorable. These advances in qubit characterization and validation methods underpin the considerable prospects for silicon as a qubit platform for fault-tolerant quantum computation.
\end{abstract}

\maketitle

\begin{figure}
 \includegraphics[width=8.5cm]{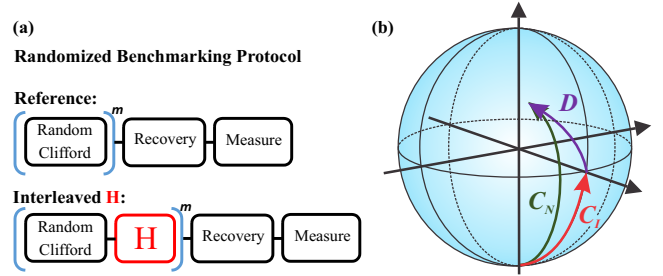}
 \caption{\textbf{a$)$} Randomized benchmarking consists of applying multiple sequences of random Clifford gates, a final recovery Clifford to ensure that each sequence ends with the qubit in an eigenstate, and reading out the qubit state. In interleaved randomized benchmarking, an additional test-Clifford gate is inserted in between the random Cliffords.  \textbf{b$)$} Bloch sphere representation for the breakdown of a general noisy operation $\cC_N$ into an ideal $\cC_I$ rotation followed by a noise operation $\cD$.}
  \label{fig:Background}
\end{figure}

Randomized benchmarking experiments~\cite{knill2008randomized,magesan2011scalable} quantify the accuracy of quantum gates by estimating the average decay in control fidelity as a function of the number of operations applied to a qubit. Benchmarking enjoys several advantages over the traditional methods of characterizing gate fidelity that involve quantum process tomography~\cite{chuang1997prescription,poyatos1997complete}, namely that it is insensitive to state preparation and measurement (SPAM) errors, and scales efficiently with the system size. As such, benchmarking protocols (see Figure~\ref{fig:Background}) have become a standard against which different qubit technologies and architectures are compared. Benchmarking experiments have been performed in many different technologies, including trapped ions~\cite{knill2008randomized,gaebler2012randomized,harty2014high}, superconducting qubits~\cite{chow2009randomized,magesan2012efficient,barends2014logic}, nuclear magnetic resonance architectures~\cite{ryan2009randomized}, nitrogen-vacancy centers in diamond~\cite{dolde2014high}, semiconductor quantum dots in silicon~\cite{veldhorst2014addressable}, and phosphorous atoms in silicon~\cite{muhonen2014quantifying}. Most experiments are fitted using an exponential fidelity decay, which is in line with original theoretical predictions~\cite{knill2008randomized,emerson2005scalable}, and consistent with the assumption of weak correlation between noise on the gates that is important for accurate fidelity estimates. 

When the assumptions of randomized benchmarking are violated, there is no guarantee of observing the characteristic exponential decay curves determined by the average fidelity. This has been noted before in NMR experiments due to spatial inhomogeneity across the sample~\cite{ryan2009randomized} as well as in numerical simulations~\cite{epstein2014} of $1/f$ noise. Recent experimental results in spin-based silicon metal-oxide-semiconductor (Si-MOS) quantum dot qubits~\cite{veldhorst2014addressable} have also shown non-exponential fidelity decay, and here we directly apply our theoretical modelling to these experiments, but our conclusions are widely applicable. 

Here we argue that non-exponential fidelity decay is indeed indicative of a dephasing-limited decay caused by non-Markovian noise. We first propose a numerical simulation method that incorporates time-dependent effects, primarily a drift in frequency detuning. This detuning drift and other time-dependent low-frequency noise sources lead to decay curves that are effectively integrated over an ensemble of experimental results, each with slightly different ``instantaneous'' average fidelities, i.e., fidelities that are approximately stable over the course of a single benchmarking run, but that drift over the course of the entire sequence of experiments. These simulations show good qualitative agreement with the observed non-exponential decay from the experiments on isotopically-purified silicon quantum dot qubits~\cite{veldhorst2014addressable}. We then give a more quantitative analysis that compares two very simple models that both give good fits to the data: the first is a simplified version of the drift model that postulates that each experimental run has one of only two possible average fidelities; the second model attributes the non-exponential decay to fluctuating SPAM errors. Both of these models have only one additional parameter over the standard benchmarking model, but our quantitative likelihood analysis shows that the simplified drift model is much more probable. 

The conclusion of this analysis for the SiMOS quantum-dot qubit is that, while the total average fidelity over a long series of benchmarking runs is 99.6\% \cite{veldhorst2014addressable}, the instantaneous fidelity can be as high as 99.9\% or more when naturally fluctuating environmental noise sources and system calibrations are most favorable. Achieving such high fidelities for single-qubit gate operations gives optimism for exceeding the demanding error thresholds for fault-tolerant quantum computation.

\section{Benchmarking Review}

The standard randomized benchmarking procedure involves subjecting a quantum system to long sequences of randomly sampled Clifford gates followed by an inversion step and a measurement, as depicted in Figure~\ref{fig:Background}. The unitary operations of the Clifford group $\Cliff$ are those that map the set of Pauli operators to itself under conjugation. They are a discrete set of gates that exactly reproduce the uniform average gate fidelity, averaged over the set of all input pure states~\cite{Dankert2009}. An alternate version known as interleaved benchmarking~\cite{magesan2012efficient} inserts a systematic application of a given gate, such as the $H$ gate shown in Figure~\ref{fig:Background}. The difference from the reference sequence gives information about the specific average gate fidelity of the given gate, rather than the average fidelity additionally averaged over the ensemble of gates.

Consider a general noise process $\cD$, depicted in Figure~\ref{fig:Background}, which represents the deviation of a noisy Clifford gate $\cC_N$ from an ideal unitary Clifford operation $\cC_I$:
\[\cC_N = \cD\cC_I \,.\]
We note that the above equation uses the formalism of completely positive maps~\cite{Nielsen2000}, and the multiplication corresponds to composition of maps.  The standard approach to randomized benchmarking makes the assumption that $\cD$ does not depend on the choice of $\cC_I$ or other details such as time, but our simulations and of course real experiments will include such a dependence. 

The fundamental result of randomized benchmarking~\cite{magesan2011scalable} is that for sufficiently well-behaved noise the observed fidelities only depend on the average error operation $\cE_\cD$ averaged over the Clifford group $\Cliff$ given by
\begin{equation*}
\cE_\cD = \frac{1}{|\Cliff|}\sum_{\cC_I\in\Cliff}\cC^{\vphantom{\dagger}}_I\cD\cC_I^{-1} \,,
\end{equation*}
as well as any SPAM errors present in the system. Furthermore, standard tools from representation theory reduce this average error operation to one that is nearly independent of $\cD$, and is characterized by just a single parameter $p$. In particular, it is a depolarizing channel $\cE$ with $p = p(\cD)$ being the polarization parameter (i.e., the probability of the information remaining uncorrupted as it passes through the channel). For a $d$-dimensional quantum system, the action of the depolarizing channel is given by
\[\mathcal{E}(\rho) = p\rho + (1-p)\frac{\mathbb{1}}{d}\,,\]
and the polarization parameter is related to the noisy deviation $\cD$ by the average gate fidelity $\bar{\cF}_{\text{avg}}(\cD)$ according to~\cite{magesan2011scalable} 
\begin{equation}\label{E:Favg}
\bar{\cF}_{\text{avg}}(\cD) = \int \mathrm{d}\psi \langle \psi |\cD(|\psi\rangle\langle\psi|)|\psi\rangle = p + \frac{1-p}{d}\,,
\end{equation} 
where the integral is a uniform average over all pure states. 

For a randomized benchmarking sequence comprised of $m+1$ total Clifford gates (including the +1 for the recovery operation), the average sequence fidelity is given by~\cite{magesan2011scalable} 
\begin{equation}\label{eq:fidelity}
\bar{F}_m = Ap^m+B \,.
\end{equation}
Here the parameters $A$ and $B$ quantify the SPAM errors and are given by~\cite{magesan2011scalable} 
\[A = \Tr\bigl[E \cD(\rho - \mathbb{1}/d)\bigr] \quad , \quad B=\Tr\bigl[E \cD(\mathbb{1}/d)\bigr] \,,\]
and $\rho$ and $E$ are the noisy state preparations and measurements implemented instead of the ideal desired states and measurements.

A typical benchmarking experiment proceeds by estimating $\bar{F}_m$ for several values of $m$ and fitting to the model in Eq.~\ref{eq:fidelity} to extract the $p$, $A$, and $B$ fit parameters, and then using Eq.~\ref{E:Favg} to report an ensemble average of the average gate fidelities $\bar{\cF}_{\text{avg}}$ of the gates.

This derivation of Eq.~\ref{eq:fidelity} assumes certain features about the noise, namely that it has negligible time and gate dependence, and that non-Markovian effects are not present at timescales on the order of the gate time. The limits to the validity of this assumption have been probed before~\cite{epstein2014, wallman2014randomized, Granade2015}, and in particular it was noted via numerical simulations by Epstein \textit{et al.}~\cite{epstein2014} that the exponential model of fidelity decay no longer holds in the presence of $1/f$ noise, resulting in a noise floor to the accuracy of the benchmarking experiment.

\section{Non-exponential Fidelity Decay}
A clear deviation from the fidelity decay predicted by Eq.~\ref{eq:fidelity} has been observed in a silicon quantum dot qubit~\cite{veldhorst2014addressable}. In order to understand the possible origin of this deviation, we have used the qubit characteristics to numerically simulate randomized benchmarking with a realistic noise model. In the experiment, the qubit is defined by the spin state of a single electron. A magnetic field $B_0 = 1.4$ T is applied to create a Zeeman splitting and the qubit is operated using electron spin resonance (ESR) techniques by applying an AC magnetic field with frequency $\omega_0 = \frac{g\mu_BB_0}{\hbar}$. A Rabi $\pi$-pulse is realized in $\tau_{op} = 1.6$ \textmu s and using a Ramsey sequence the dephasing time $T_2^* = 120$ \textmu s has been obtained~\cite{veldhorst2014addressable}. In between consecutive pulses, a waiting time $\tau_w = 0.5$ \textmu s has to be incorporated, due to the operation of the analog microwave source. 

\begin{figure}
 \includegraphics[width=\columnwidth, trim= 65 26 82 59, clip=true]{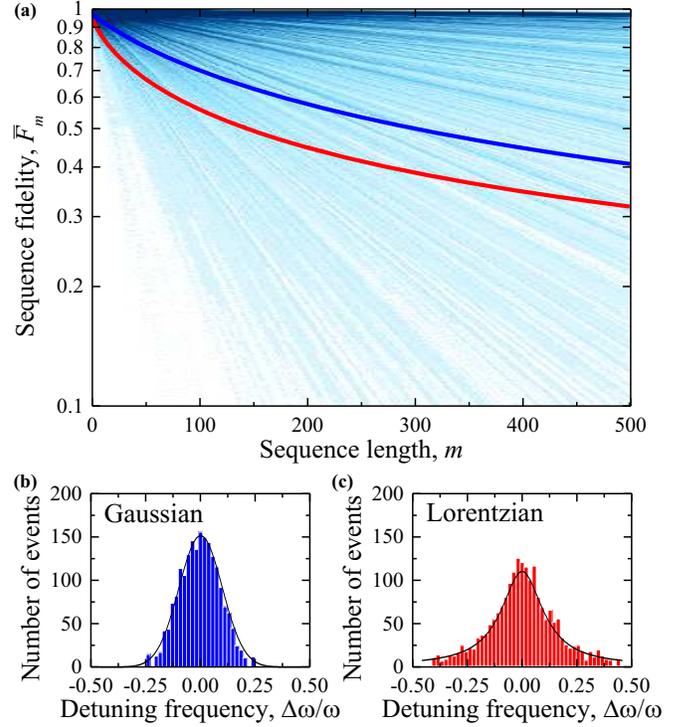}
 \caption{{\bf a$)$} Sequence fidelity as a function of sequence length $m$, with the qubit subject to Gaussian distributed $T_2^*$ associated noise. Each light blue line represents a fidelity decay for one particular value of detuning $\Delta\omega$. The linear decay on the logarithmic scale illustrates that these individual traces are indeed exponential while the ensemble average (bold blue) is non-exponential. The bold red line is the ensemble average for a Lorentzian distributed noise. {\bf b$)$} Gaussian distributed detuning frequencies and {\bf c$)$} Lorentzian distributed detuning frequencies associated with individual traces.}
 \label{fig:GaussianSpecTrace}
\end{figure}

The set of Clifford gates is generated using the set $[\pm X,\pm \frac{1}{2}X,\pm Y,\pm \frac{1}{2}Y]$ that are realized using Rabi pulses, and the identity simulated with a waiting time equal to a $\pi$-pulse. The noise processes which determine $T_2^*$ can be modelled as a random walk of the detuning $\Delta\omega$ away from the ideal operation frequency $\omega_0$, over timescales greater than a single run of a random Clifford sequence. In order to simulate an ensemble of results, the $\Delta\omega$ term is selected randomly from a Gaussian distribution of normalized variance:  
\begin{equation}\label{eq:Distribution}
\sigma_{\text{op}} = \frac{\tau_{\text{op}}}{2\pi\sqrt{2\ln(2)}T_2^*}\,.
\nonumber
\end{equation} 
Using this distribution, we have numerically simulated benchmarking experiments and the results are shown in Fig. \ref{fig:GaussianSpecTrace}. The individual traces correspond to a given detuning $\Delta\omega$ and result in the ``instantaneous'' fidelity of the qubit. While the individual traces are decaying exponentially, the averaged fidelity (bold blue) obtained from the Gaussian ensemble is clearly non-exponential. We have also included the case of a Lorentzian distribution of detunings (red), resulting in a non-exponential decay as well. In the simulation, the only error source is dephasing, whereas in the experiment, other errors might be present such as pulse-errors. Inclusion of such errors will still result in non-exponential decays, provided dephasing is a significant source of error. We note that in the experiment, Ramsey sequences have been performed in between sequences to recalibrate the resonance frequency of the qubit and to compensate drifts due to, for example, the superconducting magnet. These drifts, in combination with errors in setting the resonance frequency, result in an apparent $T_2^*$ in the randomized benchmarking experiment that is dependant upon the duration of the data acquisition, causing a faster decay of the ensemble averaged sequence fidelity. 

As low-frequency drift of the qubit resonance frequency can lead to non-exponential fidelity decay, we hypothesize that some ensemble of experiments with varying decay rates is the correct explanation for the non-exponential behaviour of the experimental benchmarking data~\cite{veldhorst2014addressable}. To support this hypothesis, we use the Akaike information criterion to show that a simple model allowing for differing fidelity rates better explains the data than an alternative explanation that assumes fluctuating SPAM errors in the standard (zeroth order) model. 

\subsection{Eliminating the constant for a single-qubit randomized benchmarking model}
The parameters $A$ and $B$ in Eq.~\ref{eq:fidelity} are nuisance parameters that do not convey information about the desired fidelity. Eliminating one of these parameters, in this case $B$, will further constrain the zero order model and allows deviations to be more clearly identifiable. A further advantage of removing the parameter $B$ is to allow fitting of a linear function on a log-linear plot.   
In Ref.~\cite{veldhorst2014addressable} the randomized benchmarking protocol was modified to eliminate $B$ from the zero order model. We first provide a theoretical justification for this approach, which we note applies only to qubits ($d=2$), and demonstrate that the resulting data highlight the deviation of the measured data from the expected exponential decay model.

Recall that the zero-order model fits the average fidelity of a gate sequence to a simple formula as follows~\cite{magesan2011scalable}:
\begin{equation}
	\Fup =A^\uparrow p^m+B^\uparrow \,,
\end{equation}
where the qubit is initialized as $\proj{\uparrow}$, the final gate in the random benchmarking sequence is chosen to return the state to $\proj{\uparrow}$, and $\Fup$ is the survival probability of this state. To eliminate the constant $B^\uparrow$ from this sequence it is only necessary to perform similar randomized sequences, save that the final $(m+1)^{\text{th}}$ gate is set to change the state to $\proj{\downarrow}$.  For these runs, we consider the survival probability for yielding the measurement outcome $E^\downarrow$, where in the ideal case the final state  $\rho = E^\downarrow = \proj{\downarrow}$. This is the survival probability for each run $\Fdown$. Under the same assumptions we have
\begin{equation}
	\Fdown=A^\downarrow p^m+B^\downarrow \,.
\end{equation}

Combining these two equations by defining $\tilde{F}_m \equiv \Fup - (1-\Fdown)$, we have:
\begin{equation}\label{eq:ftilde}
  \tilde{F}_m = \tilde{A}p^m+(B^\uparrow+B^\downarrow) -1 \,,
\end{equation}
where $\tilde{A} = A^\uparrow + A^\downarrow$.

Recall that $B^\uparrow =\Tr\bigl[E^\uparrow \cD(\mathbb{1}/d)\bigr]$, where $\cD$ is the average noise operator. For the $\Fdown$ runs, the derivation is identical, apart from the final change to the $\proj{\downarrow}$ state, so we have  
$B^\downarrow = \Tr\bigl[E^\downarrow \cD(\mathbb{1}/d)\bigr]$. 
Noting that $E^\uparrow + E^\downarrow = \mathbb{1}$ for qubits ($d=2$) and that $\cD$ is trace-preserving and assuming the error on the final $X$ gate is negligible, $B^\downarrow$ can be re-expressed as follows:
\begin{align}
B^\downarrow &=\Tr\bigl[E^\downarrow \cD(\mathbb{1}/2)\bigr]
 =\Tr\bigl[(\mathbb{1}-E^\uparrow)\cD(\mathbb{1}/2)\bigr]\nonumber\\
 &=\Tr\bigl[\cD(\mathbb{1}/2)\bigr] - \Tr\bigl[E^\uparrow \cD(\mathbb{1}/2)\bigr]
=1-B^\uparrow \,.  \label{eq:b0}
\end{align}
Therefore by subtracting the average results of the data-set $(1-\Fdown)$ from the average results of the data-set $\Fup$ we can obtain a data set that is distributed according to the model 
\begin{align}
\tilde{F}_m &= \tilde{A}p^m \,\label{eq:noB}
\end{align}
under the standard benchmarking assumptions on the noise. 

The data from Ref.~\cite{veldhorst2014addressable} consist of 8 data sets (one reference set and 7 interleaved sets). For each data set, gate sequences of the lengths \{2, 3, 5, 8, 13, 21, 30, 40, 50, 70, 100, 150\} were measured, where for each gate sequence length ($m$) the randomized protocol was carried out 500 times and randomly distributed over $\Fup$ and $\Fdown$. This amount of randomization is at least an order of magnitude more than in previous experiments~\cite{knill2008randomized, gaebler2012randomized, magesan2012efficient, chow2009randomized, barends2014logic, ryan2009randomized, dolde2014high, muhonen2014quantifying}. Each randomized protocol was performed 50 times in order to estimate the survival probability.

\subsection{Analysis of experimental data}

To quantify the quality of our fits, we require estimates of the variance in the data of Ref.~\cite{veldhorst2014addressable}.  The observed variance of the data matched to within 5-40\% of the theoretical upper bounds derived in Ref.~\cite{wallman2014randomized} when the gate length was shorter than 20 (so that the $m (1-\bar{\cF}_\text{avg}) \ll1$ assumption discussed in that reference was satisfied). Accordingly, the observed experimental variance was used as a reliable estimate of the actual variance of the distribution. It should be noted that the observed variance actually decreased for gate lengths of 100 or greater. One explanation for this unexpected behaviour is that some of the sequences become saturated to something close to a completely mixed state before reaching those sequence lengths.

\begin{figure}
 \includegraphics[width=\columnwidth, trim= 62 28 99 58, clip=true]{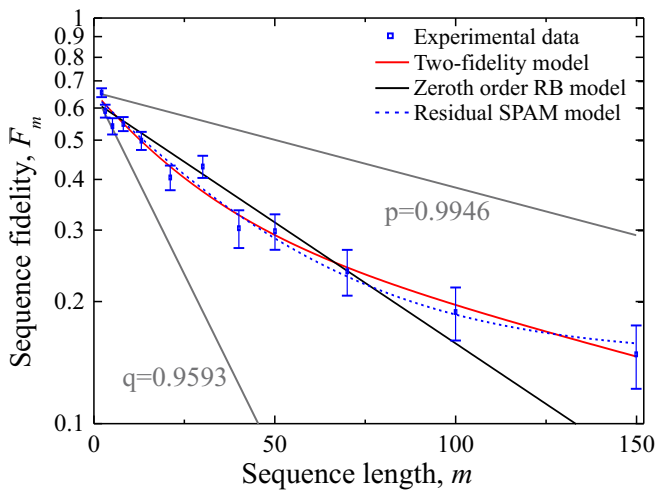}
 \caption{Semi-log plot of $F_m^\uparrow - (1-F_m^\downarrow)$ for the reference sequence of randomized benchmarking on a silicon quantum dot qubit~\cite{veldhorst2014addressable}. Both the two fidelity model and a single fidelity model including residual SPAM can fit the data, but for the single fidelity model an unreasonably large SPAM has to be included.} 
    \label{fig:fittedPlot}
\end{figure}

Figure \ref{fig:fittedPlot} shows the data from the reference dataset plotted on a semi-log plot. The confidence bounds are 95\% and the data is clearly non-linear (i.e.\ the decay is not a simple exponential). Similar deviation from the linear fit was noted in each of the data sets, with the best-fit linear model consistently underestimating $\bar{F}_m$ for $m \ge100$. 

Two possible explanations are considered. First, it may not be possible to entirely eliminate the constant term ($B$) due to a violation of one of the assumptions in the above derivation.  A second explanation is that low-frequency noise leads to detuning, and hence time-dependent errors on the gates in some of the experiments.  The first, which we denote the \textit{residual SPAM model}, can be modelled by reverting to a formula of the form $\tilde{F}_m = \tilde{A}p^m + \tilde{B}$, where now $\tilde{B}$ represents residual SPAM errors that were not eliminated under the assumptions that led to the derivation of Eq.~\ref{eq:noB}. We consider the simplest possible model for the second explanation -- the \textit{two fidelity model} -- by fitting the fidelity decay to a formula of the form $\tilde{F}_m = \tilde{A}p^m + \tilde{A}q^m$. This represents an attempt to model the data by simplifying the ensemble of experiments by reducing them to just two different \emph{equally weighted sequence behaviours:}  one with a high-fidelity rate (related by the usual measure to $p$) and one with a lower fidelity rate (similarly related to $q$).  This model has fewer parameters than the Gaussian or Lorentzian drift models, and is much easier to fit. In this interpretation, we have successfully eliminated the $B$ parameter as per Eq.~\ref{eq:noB}, but time variation gives us the two different polarization parameters $p$ and $q$, with the decay rate for each sequence sampled randomly with equal probability.  As can be seen in Figure~\ref{fig:fittedPlot}, both models fit the data substantially better than the simple exponential of the zero order model. 

Although the residual SPAM model produces a good fit to the experimental data, it does so with the equivalent of an unusually large SPAM parameter $\tilde{B}$ of around 0.14, corresponding to a $B^\uparrow$ of 0.57.  This represents in the theoretical model a very large bias in the expectation value of the spin-up measurement on the completely mixed state away from 0.5, which is not observed in the experiment.  This suggests that this model may not be best for explaining the observed data.

\begin{table}[t]
\begin{tabular}[c]{lccc}
 \hline
 &\multicolumn{2}{c}{Akaike Information Criteria}\\
 Dataset$\quad$ & $\quad \tilde{A}p^m + \tilde{B}\quad$ &  $\quad \tilde{A}p^m + \tilde{A}q^m\quad$ & $\quad$Comparison \rule[-2ex]{0pt}{5ex} \\
 \hline
 Ref & -16.93 & -25.29 & \phantom{00}65.44\\
 I	& -46.19 & -57.12 & \phantom{0}238.10\\
 X & -54.52 & -59.99 & \phantom{00}15.43\\
 X/2 & -62.89 & -63.79 & \phantom{000}1.56\\
 -X/2 & -57.77 & -64.34 & \phantom{00}26.69\\
 Y & -36.06 & -50.43 & \phantom{00}1317\\
 Y/2 & -36.04 & -46.39 & \phantom{0}172.0\\
 -Y/2 & -46.37 & -63.32 & \phantom{00}4815\\
 \hline
 \end{tabular}
 \caption[Akaike Information Criterion]{Akaike information criterion for standard and interleaved randomized benchmarking. The comparison column specifies how many times as probable is the $\tilde{A}p^m + \tilde{A}q^m$ model to minimize information loss as compared to the $\tilde{A}p^m + \tilde{B}$ model. }
 \label{table:aic}
 \end{table}

To compare these two models quantitatively, it is possible to calculate the log likelihood and Akaike information criterion\cite{Akaike1974} for the two models.  Because we don't have the actual distribution of the test statistic, we make the assumption that the samples contained in the underlying data are independent and the Gaussian distributed limit is appropriate.  This assumption is well-justified as we have a large number of independent data sets.  The distribution $\tilde F_m$ can therefore be approximated by a Gaussian distribution with a variance estimated by the observed variance at each gate length. The log likelihood of the observed data, given each of the two models, can then be calculated using standard methods.

Table~\ref{table:aic} shows the calculated Akaike information criterion for each of the experimental datasets. As can be seen, the two fidelity model better explains the data, significantly so on all but one of the datasets. Although such a model is a simplified version of the drift model, the fact that it fits the data well and is physically motivated supports its adoption as the most likely explanation of the non-exponential curve seen in the data. 

\subsection{Interpreting the two fidelity model}

 \begin{figure}[t]
  \includegraphics[width=\columnwidth, trim= 75 20 92 58, clip=true]{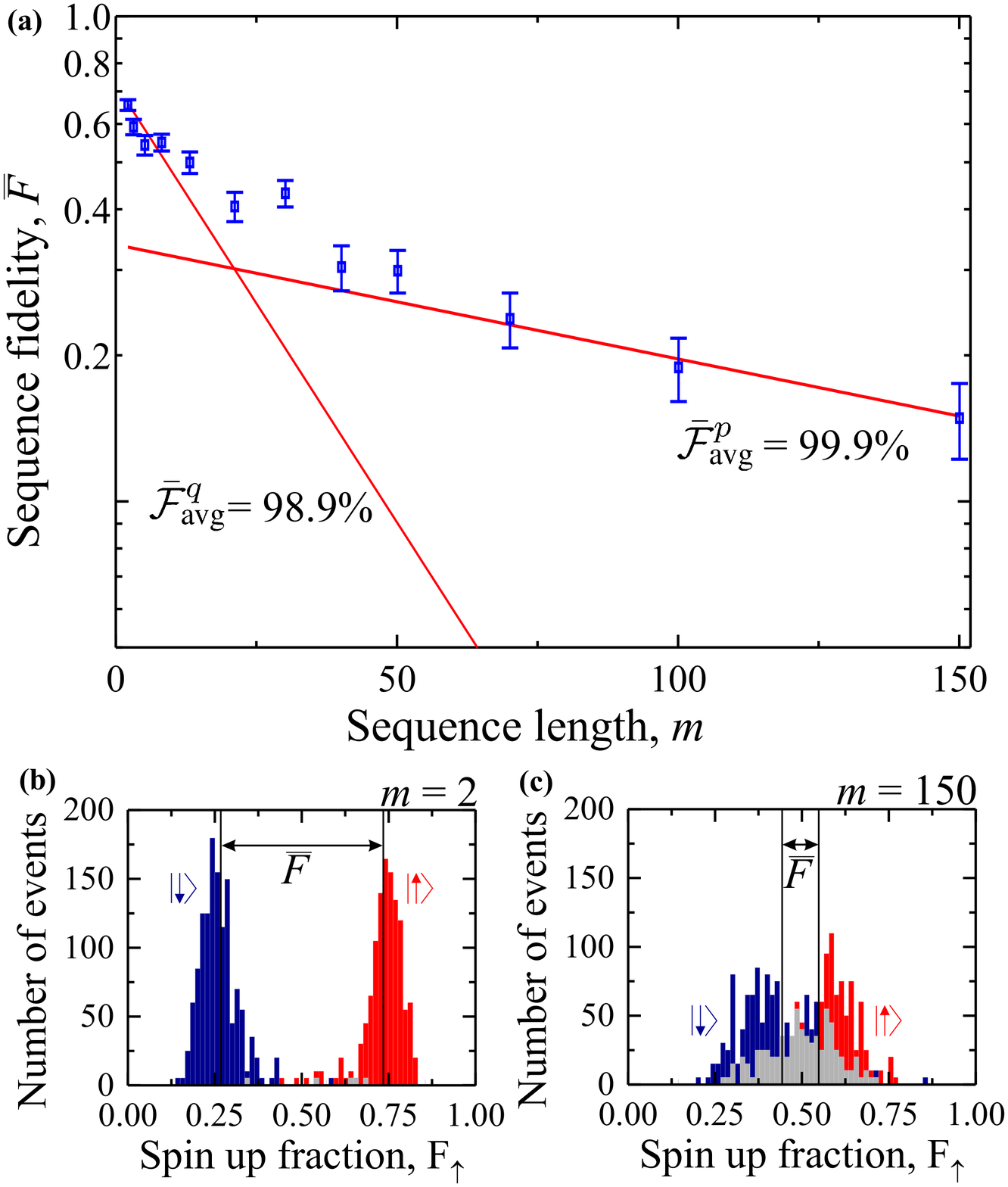}
  \caption{\textbf{$($a$)$} Reference sequence of randomized benchmarking on a silicon quantum dot qubit~\cite{veldhorst2014addressable}. The separate fidelities from the two fidelity model have been plotted to show how the initial decay is dominated by the low $q$ value, whereas the higher value of $p$ is indicative of the average decay in the longer lived high fidelity regime. Histogram of spin-up $\ket\uparrow$ and spin-down $\ket\downarrow$ corresponding with data point $m = 2$ \textbf{$($b$)$} and $m = 150 $ \textbf{$($c$)$}. Results with expected spin up outcome are shown in red while blue represents data with expected spin down result. The grey regions illustrate the overlapping areas.} 
  \label{fig:Fittings&Hist}
\end{figure}

Since the two fidelity model is the quantitatively preferred model, a natural question arises: how should we interpret the model parameters? The obvious interpretation of the two parameters $p$ and $q$ is as presented in table~\ref{table:aic2}; that their difference represents the characteristic spread of the actual underlying ensemble of fidelities from which the benchmarking data are sampled. Such an interpretation is natural and compelling, however it remains an open problem to quantify such a connection more carefully. In particular, it would be interesting to give a direct connection to a more general drift model, since these are easier to interpret physically, but much harder to fit and analyze statistically. 

By considering the non-exponential decay manifesting as the average over an ensemble of results, the fidelity can be considered to be operating under two regimes as depicted in Figure~\ref{fig:Fittings&Hist}a. Firstly, dominating the observed fidelity decay at low $m$, there is rapid and short-lived decay due to traces of large detuning $\Delta\omega$. Secondly, for large $m$, there is an approximately exponential tail due to long-lived traces of smaller detuning. The second regime can be approximated as exponential due to the larger detuning frequencies becoming negligible as $m$ increases. In Figure~\ref{fig:Fittings&Hist}a, each of the data points are an average over 25,000 experimental repetitions as presented by the two accompanying histograms (Figure~\ref{fig:Fittings&Hist}b for $m = 2$ and Figure~\ref{fig:Fittings&Hist}c for $m = 150$). Each histogram separately shows the measured probability, averaged over $50$ repetitions, for the spin-up and spin-down observables as expected at the end of a noiseless version of the applied random sequence. From the second regime of Figure~\ref{fig:Fittings&Hist}a we can see that many experiments within the ensemble of measurements have an instantaneous fidelity at least as high as 99.9\%.

\begin{table}
\begin{tabular}[c]{ccclllc}
 \hline
 
 Dataset&$p$&$q$&$\bar \cF_{\text{avg}}^{p/p}$&$\bar \cF_{\text{avg}}^{q/q}$&$\bar \cF_{\text{avg}}^{q/p}$&Uncertainty
  \rule[-2ex]{0pt}{5ex} \\ \hline
 Ref & 0.995 & 0.959 & 99.9\%&98.9\%&$-$&0.06\% \\
 I	& 0.993 & 0.946 & 99.9\%&99.6\%&98.7\%&0.3\%\\
 X & 0.993 & 0.952 & 99.9\%&99.8\%& 98.9\% & 0.2\% \\
 X/2 & 0.993 & 0.947 & 99.9\%&99.7\%&98.7\%&0.2\% \\
 -X/2 & 0.991 & 0.947 & 99.9\%&99.7\% &98.7\%&0.2\%\\
 Y &0.993 & 0.964& 99.9\%&99.9\%& 99.1\%&0.3\%\\
 Y/2 & 0.991 & 0.952 & 99.9\%&99.8\%&98.9\%&0.2\%\\
 -Y/2 & 0.990 & 0.911 & 99.9\%&98.7\% &97.8\%&0.2\%\\
 \hline
 \end{tabular}
 \caption[Gate fidelity estimates ($\bar \cF_\text{avg}$) for each of the datasets.]{Calculated $p$ and $q$ values for the two fidelity model. The gate fidelity estimates ($\bar \cF_\text{avg}$) reported for the reference run are the high $(p)$ gate fidelity estimate and low $(q)$ gate fidelity respectively. For the interleaved models, three comparative estimates are reported, the first  $\bar \cF_\text{avg}^{p/p}$ is calculated by comparing the higher fidelity of the interleaved run with the higher fidelity from the reference run, the second $\bar \cF_\text{avg}^{q/q}$ by comparing the lower fidelity of the interleaved run with the lower fidelity of the reference run, and the third $\bar \cF_\text{avg}^{q/p}$ (which represents the worst possible method of calculating this) compares the low interleaved run with the higher reference run. The error for the reference set is calculated directly from the data fit. For the interleaved runs the formulas provided in Ref.~\cite{magesan2012efficient} were used to determine the likely error margins.} 
 \label{table:aic2}
 \end{table}

\section{Conclusions}

We have analyzed the non-exponential decay in randomized benchmarking experiments on Si-MOS quantum dot qubits, and found that the most plausible explanation of this decay is drift in detuning frequencies. Our simulation of temporal integration over a spectrum of time-dependent detuning frequencies qualitatively reproduces the observed fidelity decay of previously conducted experiments~\cite{veldhorst2014addressable}. In addition, we have quantitatively ruled out a competing model by showing agreement of a simplified ensemble (the two fidelities model) that is much more probable. This yields confidence that detuning drift is the correct explanation for the origin of such a non-exponential fidelity decay. 

Fitting the randomized benchmarking data with a two-fidelity model demonstrates that silicon MOS quantum dot qubits already exhibit an ``instantaneous'' control fidelity of 99.9\%. We anticipate that this value is the relevant fidelity for quantifying the achievable performance of these gates for quantum computation, since improvements in the readout fidelity and use of a fast Ramsey protocol to calibrate the resonance frequency for each experiment~\cite{Shulman2014} could result in an ensemble fidelity that matches the best instantaneous fidelity which is ultimately defined by fixed errors.

These results raise several intriguing questions. The first is to quantitatively link the simple and easy to analyze two fidelity model to the Gaussian or Lorentzian drift models. Alternatively, directly fitting a drift ensemble to the data would give a better picture of the source of the non-exponential fidelity decay, but this approach risks overfitting, and is already difficult for the simple case of Gaussian-distributed detunings. 

Finally, there is at least one other natural competing explanation for the non-exponential decay. It might be the case that long benchmarking sequences saturate the exponential decay rates and have slower decay on very long timescales. If this were the case, then fitting to sequences that were ``too long'' would certainly bias one toward seeing non-exponential decay and reporting fidelities that were higher than warranted by the analysis. Therefore, deriving stopping criteria for the maximum sequence length and deriving tests that rule out this alternate explanation is a further important open question for future work.

\acknowledgments 

We thank Chris Ferrie for helpful discussions. The authors acknowledge support from the Australian Research Council (CQC2T - CE11E0001017 and EQuS - CE11001013), the NSW Node of the Australian National Fabrication Facility, the US Army Research Office (W911NF-13-1-0024, W911NF-14-1-0098, W911NF-14-1-0103 and W911NF-14-1-0133), and by iARPA via the MQCO program. M.V. also acknowledges support from the Netherlands Organization for Scientific Research (NWO) through a Rubicon Grant. S.T.F. also acknowledges support from an ARC Future Fellowship (FT130101744).



\begin{thebibliography}{99}

\bibitem{knill2008randomized} E.\ Knill \textit{et al.}, Randomized benchmarking of quantum gates, Phys.\ Rev.\ A \textbf{77}, 012307 (2008).

\bibitem{magesan2011scalable} E.\ Magesan, J.M.\ Gambetta, and J.\ Emerson, Scalable and robust randomized benchmarking of quantum processes, Phys.\ Rev.\ Lett.\ \textbf{106}, 180504 (2011).

\bibitem{chuang1997prescription} I.\ Chuang and M.\ Nielsen, Prescription for experimental determination of the dynamics of a quantum black box, J.\ Mod.\ Opt.\ \textbf{44}, 2455 (1997).

\bibitem{poyatos1997complete} J.F.\ Poyatos, J.I.\ Cirac, and P.\ Zoller, Complete characterization of a quantum process: the two-bit quantum gate, Phys.\ Rev.\ Lett.\ \textbf{78}, 390 (1997).

\bibitem{gaebler2012randomized} J.P.\ Gaebler \textit{et al.}, Randomized benchmarking of multiqubit gates, Phys.\ Rev.\ Lett.\ \textbf{108}, 260503 (2012).

\bibitem{harty2014high} T.P.\ Harty, \textit{et al.}, High-Fidelity Preparation, Gates, Memory, and Readout of a Trapped-Ion Quantum Bit, Phys.\ Rev.\ Lett.\ \textbf{113}, 22, 220501 (2014).

\bibitem{chow2009randomized} J.M.\ Chow \textit{et al.}, Randomized benchmarking and process tomography for gate errors in a solid-state qubit, Phys.\ Rev.\ Lett.\ \textbf{102}, 090502 (2009).

\bibitem{magesan2012efficient} E.\ Magesan \textit{et al.}, Efficient measurement of quantum gate error by interleaved randomized benchmarking, Phys.\ Rev.\ Lett.\ \textbf{109}, 080505 (2012).

\bibitem{barends2014logic} R.\ Barends \textit{et al.}, Superconducting quantum circuits at the surface code threshold for fault tolerance, Nature \textbf{508}, 500-503 (2014).

\bibitem{ryan2009randomized} C.A.\ Ryan, M.\ Laforest and R.\ Laflamme, Randomized benchmarking of single-and multi-qubit control in liquid-state NMR quantum information processing, New J.\ Phys.\ \textbf{11}, 013034 (2009).

\bibitem{dolde2014high} F.\ Dolde \textit{et al.}, High-fidelity spin entanglement using optimal control, Nat.\ Comm.\ \textbf{5}, 3371 (2014).

\bibitem{veldhorst2014addressable} M.\ Veldhorst \textit{et al.}, An addressable quantum dot qubit with fault-tolerant control-fidelity, Nat.\ Nano.\ \textbf{9}, 981 (2014).

\bibitem{muhonen2014quantifying} J.T.\ Muhonen \textit{et al.}, Quantifying the quantum gate fidelity of single-atom spin qubits in silicon by randomized benchmarking, arXiv preprint arXiv:1410.2338 (2014).

\bibitem{emerson2005scalable} J.\ Emerson, R.\ Alicki and K.\ {\.Z}yczkowski, Scalable noise estimation with random unitary operators, J.\ Opt.\ B \textbf{7}, S347 (2005).

\bibitem{epstein2014} J.M.\ Epstein, A.W.\ Cross, E.\ Magesan and J.M.\ Gambetta, Investigating the limits of randomized benchmarking protocols, Phys.\ Rev.\ A \textbf{89}, 062321 (2014).

\bibitem{Dankert2009}
C.\ Dankert, R.\ Cleve, J.\ Emerson, and E.\ Livine, Exact and approximate unitary 2-designs and their application to fidelity estimation, Phys.\ Rev.\ A \textbf{80}, 012304 (2009).

\bibitem{Nielsen2000}
M.A.\ Nielsen and I.L.\ Chuang, \textit{Quantum Computation and Quantum Information}, Cambridge University Press, (2000).

\bibitem{wallman2014randomized} J.J.\ Wallman and S.T.\ Flammia, Randomized Benchmarking with Confidence, New J.\ Phys.\ \textbf{16}, 103032 (2014).

\bibitem{Granade2015}
C.\ Granade, C.\ Ferrie, and D.G.\ Cory, Accelerated Randomized Benchmarking, New J.\ Phys.\ \textbf{17}, 013042 (2015).

\bibitem{Elzerman2004Single}  J.M.\ Elzerman \textit{et al.} Single-shot read-out of an individual electron spin in a quantum dot, Nature \textbf{430}, 431 (2004).

\bibitem{Akaike1974}
H.\ Akaike, A new look at the statistical model identification, \textit{IEEE Trans. Auto. Control}, \textbf{19}, 716 (1974).

\bibitem{Shulman2014}
M.D.\ Shulman \textit{et al.} Suppressing qubit dephasing using real-time Hamiltonian estimation, Nat.\ Comm.\ \textbf{5}, 5156 (2014).

\end{thebibliography}
\end{document}